# Hardware in the loop simulation and hierarchical predictive control of a TWPTR

M. Shahgholian[1], D. Gharavian[1*]

1: Department of Electrical and Computer Engineering, Shahid Beheshti University, Tehran, Iran
*D_Gharavian@sbu.ac.ir

**Abstract:** In this paper, we use a nonlinear hierarchical model predictive control (MPC) to stabilize the Segway robot. We also use hardware in the loop (HIL) simulation in order to model the delay response of the wheels' motor and verify the control algorithm. In Two-Wheeled Personal Transportation Robots (TWPTR), a nonlinear MPC predicts the dynamics of the system and solves the control problem efficiently, but requires the exact information of the system models. Since model uncertainties are unavoidable, the time-delay control (TDC) method is used to cancel the unknown dynamics and the unexpected disturbances. When TDC method is applied, the results show that the maximum required torque for engines is reduced by 7%. And the maximum displacement of the robot has dropped by 44% around the balance axis. In other words, robot stability has increased by 78%. This research runs the HIL simulation to implementing the control algorithms without approximation.

## 1. Introduction

With the increase in personal transportation vehicles, traffic congestion is growing worse in urban areas and is expected to aggravate over the next years. Moreover, there are some other serious problems such as lack of parking space and pollution [1]. In order to improve urban trip conditions, developing a smart and less polluting narrow vehicle can be a good idea. To achieve this idea, many intelligent robots and vehicles have been applied base on two-wheeled inverted pendulum (TWIP) models[2]. Such advantages of these narrow vehicles are the occupation of less space, the possibility of sharing a single lane with another narrow vehicle, lower emission, and flexible operation.

Over the last 50 years, the inverted pendulum has been the most popular benchmark among others in nonlinear control theory [3]. These TWIPs have two wheels mounted on both sides of a chassis, with a center of mass above the wheel axles [4]. In this case, a DC motor controls each wheel independently. The control objective of the TWIP is to perform position or velocity control of the wheels, while the stability of the pendulum occurs around the vertical position which is an unstable equilibrium point. This type of systems that have numbers of actuators fewer than the degrees of freedom to be controlled is defined as under-actuated systems. Based on the mechanical configuration, under-actuated TWIPs can be categorized into a class without input coupling (Class A) and a class with input coupling (Class B). In class A, the actuator is mounted on the wheel and in class B the actuator is mounted on the pendulum or chassis [5]. The structure of the TWPTRs are based on Class A. The TWPTR is a device that transports one person at relatively low speeds. TWPTRs have a low-speed operation and use electric power systems to move. These features make this robot be a candidate for providing short-distance trips [6].

TWPTR relies on the changes in driver's gravity center to control the vehicle movements such as starting, acceleration, deceleration, stopping and so on [7]. In other words, the initial law for riding the TWPTR states that if the center of gravity (COG) leads forward, it makes the wheels accelerate and if it leads backward, it makes the wheels slow down. Essentially, the mobility of the TWPTR is not autonomous as the traveler is involved in the control [8].

Consequently, TWPTR is driven by DC motors that are applied independently of each of the wheels, thus causing the system to be under-actuated. It is also subject to the non-holonomic constraint of no sideslip. These features make the stabilization of a TWPTR a challenging and interesting control problem. In this regard, linear controllers have been successfully used to locally stabilize the pendulum around an unstable upright position. These controllers have a limited region of stability. Nonlinear fuzzy logic and adaptive neural network controllers have been effectively used to maximize the region of stabilization. The main drawback of such methods is robustness [9]–[11]. In this paper, we propose a nonlinear MPC, where this process captures the dynamic and static interactions between input, output, and disturbance variables, while the control loop is coordinated with the calculation of optimum set points. In spite of these advantages, the success of MPC depends on the accuracy of the process model [12]. Clearly, according to simulation results, this approach is very precise and rapid compared with other nonlinear methods, but like the others, it cannot handle the unmeasurement disturbances. Since model uncertainties are unavoidable for actual systems, we use the TDC approach. TDC has been known as a robust controller for a long time. TDC has been recognized as a simple, efficient, and effective control method for various nonlinear plants. The idea of the TDC method is to use the previous dynamic information to cancel out uncertainties [13]. Therefore, in this paper, we use a two-layer or hierarchical MPC as each layer covers the disadvantage of the other, while their advantages are preserved.

On the other hand, most of the published works are based on theoretical analysis, and results are obtained by simulations. However, these controllers may not operate well in real-life systems. First, the controller design and proving stability are based on the accurate mathematical models without considering any uncertainties. Second, some of the control algorithms are too complicated to be implemented [5]. Hardware-in-the-loop (HIL) simulation represents a bridge between pure simulation and complete system construction by providing an efficient real-time and safe environment. Tests can focus on the functionality of the controller and verify all the dynamic conditions of the system [14]. In other words, the HIL simulation technique is a kind of real-time simulation where the input and output signals of the simulator represent the time-dependent values like a real process. Such a simulator allows us to test the real controller or embedded control system under different real working conditions. HIL is more reliable and credible in its results than numerical simulation, and it can also save a great deal of money and time for engineers or scientific institutes [15]. In this paper, an HIL simulation is set up for TWPTR to evaluate the real-time hierarchical MPC with regards to their accuracy, computational ability, and robustness against disturbances and erroneous system parameters.

This paper is organized as follows. The next section presents the mathematical analysis of the proposed control approach. In this section, first we discuss the dynamic equation of a TWPTR; secondly we state the MPC; then the TDC approach is expressed, and finally, the HIL simulation loop is determined. In the third section, the numerical and HIL simulation of the proposed approach is demonstrated. Finally, conclusions are presented.

## 2. Mathematical Model

Before proposing the points related to the Segway robot modeling, the reference coordinates systems that make mathematical analyzing feasible are defined [16].

Inertial coordinates system (e-frame): The origin of this coordinates system is the initial position of the robot's motion field, whose z-axis ($z^e$) is along the center of the earth. Its x-axis ($x^e$) and y-axis ($y^e$) directions can be tangent to the global surface of the earth in any direction.

The origin of the other coordinates systems is fixed on the base of TWPTR.

Navigation coordinates system (n-frame): The z-axis ($z^n$) of this coordinates system is perpendicular to the surface of the ground. Its y-axis ($y^n$) is along the direction of wheels' axle and its origin is set in its center.

Pendulum coordinates system (p-frame): Its z-axis ($z^p$) is in the direction of the robot's pendulum and attached to it across all positions. Its origin and y-axis ($y^p$) are the same as the origin and y-axis of the n-frame.

Wheel coordinates system (w-frame): Its y-axis ($y^w$) is the same as the $y^p$. The x-axis ($x^w$) and z-axis ($z^w$) directions are tangent to the plate of the robot's wheel and rotate with it. The w-frame origin is the center of the wheel.

In order to avoid unnecessary complexity of the equations, it is assumed that the torques produced by the DC motor of right ($\tau_{wR}$) and left ($\tau_{wL}$) wheels are always equal. So, the direction of the right and left wheels axes is always the same as each other. The transmission vector of the left wheel to the right wheel is fixed in all positions and conditions, and its length is equal to the axle length between these two wheels.

All these four frames are demonstrated in Fig. 1.

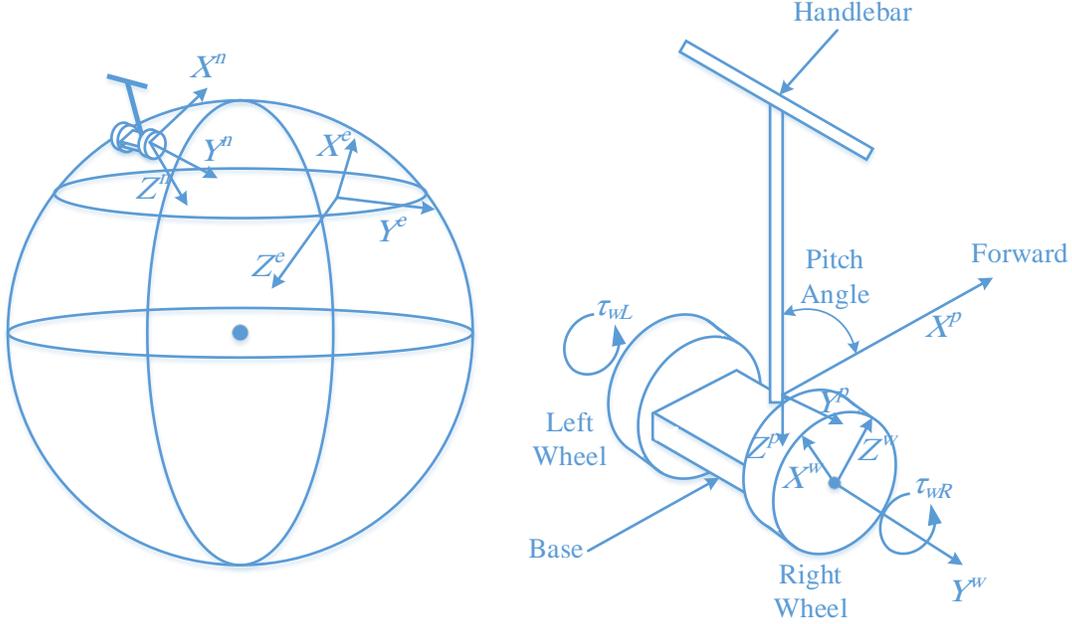

**Fig. 1:** Coordinates Systems

## 2-1.    Dynamic analysis of a TWPTR

A TWPTR is considered as Fig. 1. This robot can be divided into two parts: the wheels and the pendulum. The TWPTR's pendulum consists of a pole and motors' driver to support the body over the wheel making it balanced [17]. The p-frame of this robot can only rotate with respect to the y-axis. The motion field of the robot is assumed as a flat plane; so $x^e$ and $y^e$ can have the same directions as $x^n$ and $y^n$. Therefore, due to the equal torque of the right and left wheels, this TWPTR moves along $x^e$ or $x^n$ at all-time intervals.

- **TWPTRs' Wheel**

Since the robot moves only along the $x^n$, and the field of robot motion is flat, gravity acceleration is equal to $G^n = [0 \quad 0 \quad -g]^T$, the friction force of each wheel is $F_{tw}^n = [f_{tw}^n \quad 0 \quad 0]^T$, and the normal force is equal to $N_w^n = [0 \quad 0 \quad N_{zw}^n]^T$. In addition, the forces applied to the wheels by means of chassis are equal to $F_w^n = [f_{xw}^n \quad 0 \quad f_{zw}^n]^T$, the torque produced by each wheel's motor is $\tau_{nw}^n = [0 \quad \tau_{ynw}^n \quad 0]^T$, and the extent of friction torque caused by friction viscous between the wheels' shaft and pendulum in relation to the center of the wheel in the n-frame is $\tau_{fpw}^n = [0 \quad \tau_{fypw}^n \quad 0]^T$.

Note that all of these values are expressed in the n-frame (in all cases, the index above any sign is the indication of the reference coordinates system). The mass of each wheel is denoted by *m*. So, the forces and torques exerted to each wheel can be modeled as Fig. 2.

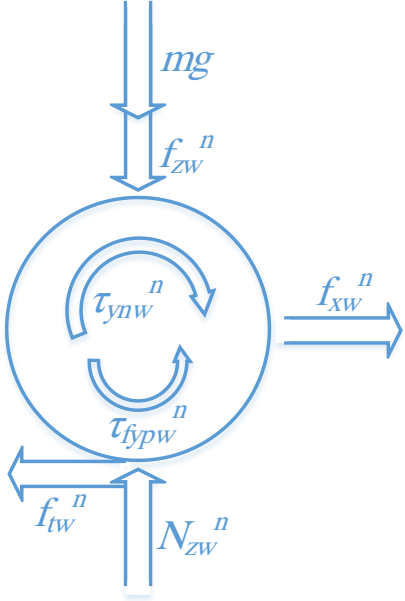

**Fig. 2:** Forces and Torques of the TWPTR's Wheel

The acceleration of each wheel is proportional to the sum of the exerted forces into the wheel minus the friction forces.

$$1. \quad m a_{ew_o}^n = F_T^n - F_{tw}^n$$

In equation 1, $a_{ew_o}^n$ is the acceleration of the w-frame origin with respect to the e-frame in the n-frame. The total force ($F_T^n$) exerted on each wheel can be calculated as equation 2.

$$2. \quad F_T^n = F_w^n + m G^n - N^n$$

Also, based on Fig. 3, the following equations can be deduced.

$$3. \quad m a_{xen}^n = f_{xw}^n - f_{tw}^n$$

4. $N_{zw}^n = mg + f_{zw}^n$

To convert the coordinates of a point in the n-frame to the e-frame, equation 5 should be used [18].

5. $P^e = C_n^e P^n + T_{en}^e$

In the equation 5, $T_{en}^e$ is the transmission vector of the n-frame to the e-frame proposed in the e-frame, $C_n^e$ is the rotation matrix of the n-frame to the e-frame, and $P^e$ and $P^n$ are the coordinates of one arbitrary point in the e-frame and the n-frame, respectively. The rotation matrix between these two frames is the unit matrix since their axes have the same directions. Only the x-axis of the transmission vector is variable($t_{xen}^n$). According to these conclusions, equation 5 can be rewritten as follows:

6. $P^e = P^n + T_{en}^n = P^n + [t_{xen}^n \quad 0 \quad 0]^T$

Because of the same direction of the e-frame and the n-frame axes, $T_{en}^e$ and $T_{en}^n$ are equal. Therefore, deriving from equation 6, the acceleration of one point in the e-frame, based on its acceleration in the n-frame can be obtained by equation 7.

7. $a^e = a^n + [\ddot{t}_{xen}^n \quad 0 \quad 0]^T$

$[\ddot{t}_{xen}^n \quad 0 \quad 0]^T$ in equation 7 is the same as $a_{ew_o}^n$ in equation 1. In other words, this acceleration is equivalent to the acceleration of the origin of the n-frame with respect to the e-frame. Since the origins of the w-frame, the p-frame, and the n-frame are fixed in relation to each other, the following relation can be written.

8. $\ddot{T}_{en}^n = \ddot{T}_{ep}^n = \ddot{T}_{ew}^n$

On the other hand, based on Fig. 3, the torque applied to each wheel can be written as equation 9 (only around the $y^w$, the torque is not zero).

$$9. \quad \tau_{ynw}^n = I_w \ddot{\varphi}_{ynw}^n - r f_{tw}^n + \tau_{fypw}^n$$

In equation 9, the moment of each wheel around $y^n$ is denoted by $I_w$. Also, according to the radius of the robot wheels, $r$, the distance traveled by the robot can be obtained by equation 10.

$$10. \quad t_{xew}^n = r \varphi_{ynw}^n$$

$\varphi_{ynw}^n$ is the pitch angle of the w-frame with respect to $y^n$. The following relation is obtained by deriving equation 10. It is the most important constraint relation that we will use in the TDC section.

$$11. \quad \dot{t}_{xen}^n = r \dot{\varphi}_{ynw}^n \rightarrow \ddot{t}_{xen}^n = r \ddot{\varphi}_{ynw}^n$$

By replacing equation 3 and equation 11, in equation 9 the dynamic model of each wheel can be rewritten as equation 12.

$$12. \quad \tau_{ynw}^n = \left( \frac{I_w}{r} + mr \right) \ddot{t}_{xen}^n - r f_x^n + \tau_{fypw}^n$$

The friction torque between w-frame and the p-frame can be calculated from the equation 13, where $b_{pw}$ is the viscous friction coefficient.

$$13. \quad \tau_{fypw}^n = b_{pw}(\dot{\varphi}_{ynw}^n - \dot{\varphi}_{ynp}^n)$$

In the above equation, $\varphi_{ynp}^n$ is the pitch angle of the p-frame with respect to $y^n$.

- **TWPTR's Pendulum**

As illustrated in Fig. 2, the pendulum is connected to the chassis of a two-wheeled robot. The wheels' torque ($\tau_{ynw}^n$) and force ($F_w^n$) are equal, which according to Newton's third law, twice of this torque is transmitted to the robot pendulum by chassis (Fig. 3).

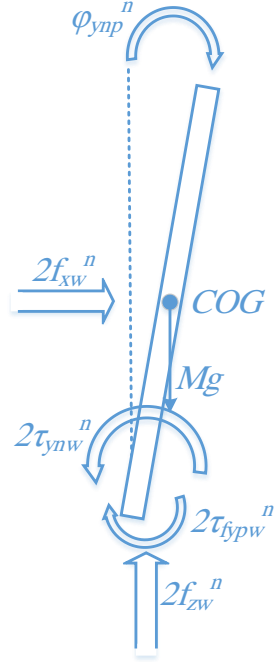

**Fig. 3:** Forces and Torques of the TWPTR's Pendulum

The coordinates of the center of gravity (COG) of the robot in the p-frame are $P_{COG}^p = [l \quad 0 \quad 0]^T$. As with equation 5, the coordinates of this point in the n-frame can be obtained through equation 14.

$$14. \quad P_{COG}^n = C_p^n P_{COG}^p + T_{np}^n$$

Since the transmission vector between these two coordinates systems is zero ($T_{np}^n = 0$), and the p-frame is only rotating around the $y^n$, equation 15 can be obtained as follows:

$$15. \quad P_{COG}^n = \begin{bmatrix} \sin \varphi_{ynp}^n & 0 & \cos \varphi_{ynp}^n \\ 0 & 0 & 0 \\ \cos \varphi_{ynp}^n & 0 & \sin \varphi_{ynp}^n \end{bmatrix} P_{COG}^p \rightarrow$$
$$x_{COG}^n = l \sin \varphi_{ynp}^n \quad , \quad z_{COG}^n = l \cos \varphi_{ynp}^n$$

In the above equation, $x_{COG}^n$ and $z_{COG}^n$, are the coordinates of the length and height of the COG along the $x^n$ and $z^n$, respectively. In order to obtain the coordinates of this point in the e-frame, we should use equation 6. The results are shown in equation 16 and equation 17.

$$16.\ x_{COG}^e = x_{COG}^n + t_{xen}^n$$

$$17.\ z_{COG}^e = z_{COG}^n$$

By deriving from the above relations, the COG acceleration is obtained easily in the e-frame.

$$18.\quad \begin{aligned} \dot{x}_{COG}^e &= \dot{\varphi}_{ynp}^n l \cos \varphi_{ynp}^n + \dot{t}_{xen}^n \rightarrow \ddot{x}_{COG}^e = \\ &+ \ddot{\varphi}_{ynp}^n l \cos \varphi_{ynp}^n - \left(\dot{\varphi}_{ynp}^n\right)^2 l \sin \varphi_{ynp}^n + \ddot{t}_{xen}^n \end{aligned}$$

$$19.\quad \begin{aligned} \dot{z}_{COG}^e &= -\dot{\varphi}_{ynp}^n l \sin \varphi_{ynp}^n \rightarrow \\ \ddot{z}_{COG}^e &= -\left(\dot{\varphi}_{ynp}^n\right)^2 l \cos \varphi_{ynp}^n - \ddot{\varphi}_{ynp}^n l \sin \varphi_{ynp}^n \end{aligned}$$

On the other hand, according to the forces exerted to the pendulum (Fig. 3), equation 20 is obtained (The mass of the pendulum is denoted by M).

$$20.\quad \begin{aligned} M\ddot{x}_{COG}^n &= 2f_{xw}^n \\ M\ddot{z}_{COG}^n &= 2f_{zw}^n + Mg \end{aligned}$$

The pendulum torque around the $y^n$ ($\tau_{ynp}^n$) can be calculated by equation 21 (The pendulum's moment around the $y^n$ is denoted by $I_p$).

$$21.\quad \begin{aligned} \tau_{ynp}^n &= I_p \ddot{\varphi}_{ynp}^n + (M\ddot{x}_{COG}^n)z_{COG}^n \\ &- (M\ddot{z}_{COG}^n)x_{COG}^n - \tau_{fypw}^n \end{aligned}$$

The values of $f_{xp}^n$, $f_{zp}^n$, $x_{COG}^n$ and $z_{COG}^n$ are incorporated from equations 15, 18, 19 and 20 in equation 21, and after the simplification, the dynamic model of the pendulum is obtained by equation 22.

$$22.\quad \begin{aligned} \tau_{ynp}^n &= \left(I_p - Ml^2\right)\ddot{\varphi}_{ynp}^n - 2\tau_{fypw}^n \\ &- Ml \cos\left(\varphi_{ynp}^n\right)\ddot{t}_{xen}^n \end{aligned}$$

By combining dynamic models of the pendulum and wheel, the equation 23 is generated.

$$23. \quad \begin{aligned} \tau_{ynw}^n &= \left(\frac{l_w}{r} + rm - \frac{rM}{2}\right) \ddot{t}_{xen}^n + \tau_{fypw}^n \\ &+ \frac{rM}{2}\left(-\ddot{\varphi}_{ynp}^n l \cos \varphi_{ynp}^n + {\dot{\varphi}_{ynp}^n}^2 l \sin \varphi_{ynp}^n\right) \end{aligned}$$

## 2-2. MPC

The angular velocity of the pendulum can be measured by a tilt sensor, a gyro sensor or an acceleration sensor [17]. In this paper, the TWPTR system is used as a gyro sensor to measure the angular velocity of the pendulum. In other words, a gyro is placed such that it measures the pendulum angular velocity around the $y^n$. Indeed, the magnitude of $\dot{\varphi}_{ynp}^n$ in the equations is determined by this sensor. The torque required to stabilize the pendulum around the upright position (keep $\varphi_{ynp}^n$ equal to zero) is also generated by the two DC motors where the shaft coupler of them is fixed at the center of each wheel.

In order to control the robot, the relationship between the acceleration of the robot chassis and the angular position of the robot pendulum should be determined. To achieve this, by equating the equation 22 and equation 23, the torques are eliminated, and the acceleration of the chassis along $x^e$ is obtained by the $\varphi_{ynp}^n$ angle, as well as its first and second derivatives. Constant coefficients in equation 24 have been defined in previous sections which depend on the robot's properties.

$$24. \ddot{t}_{xen}^n = \frac{\left(\left(-\frac{l_p}{2} + \frac{Ml^2}{2} + \frac{rMl}{2}\cos\varphi_{ynp}^n\right)\ddot{\varphi}_{ynp}^n - \frac{rMl}{2}\sin\varphi_{ynp}^n\left({\dot{\varphi}_{ynp}^n}^2\right)\right)}{\left(\frac{l_w}{r} + rm - \frac{rM}{2} - \frac{Ml}{2}\cos\varphi_{ynp}^n\right)}$$

On the other hand, using the dynamic model of the system, the behavior of the controlled variables can be predicted. In MPC, in order to compute the next state from the real state and the inputs, one must numerically solve the nonlinear differential equation 24. In order to obtain its value, a numerical method such as the Runge-Kutta method can be used[19].

The value of $\dot{\varphi}_{ynp}^n$ is determined by the gyro information in the real state (k-th sample). Using the Runge-Kutta method, the value of $\varphi_{ynp}^n$ can be predicted for the next sample time ((k+1)-th sample), while the value of $\ddot{\varphi}_{ynp}^n$ is calculated for the previous sample time ((k-1)-th sample). Ignoring the angular acceleration in the current moment, $\ddot{\varphi}_{ynp}^n(k)$ can be considered equal to $\ddot{\varphi}_{ynp}^n(k-1)$. With this calculated information, the value of TWPTR linear acceleration for the next sample time is predicted. Also, the torque required for the wheels to stabilize the pendulum constant at the next sample time is calculated by equation 23. Based on these calculations, the value of $\ddot{\varphi}_{ynp}^n$ for the present sample time is estimated. This estimation is obtained by the predicted linear acceleration and wheels' torque, and the angular position is estimated by the output of the gyro in the previous sample time (Fig. 4). A detailed discussion on MPC can be found in [12], [20].

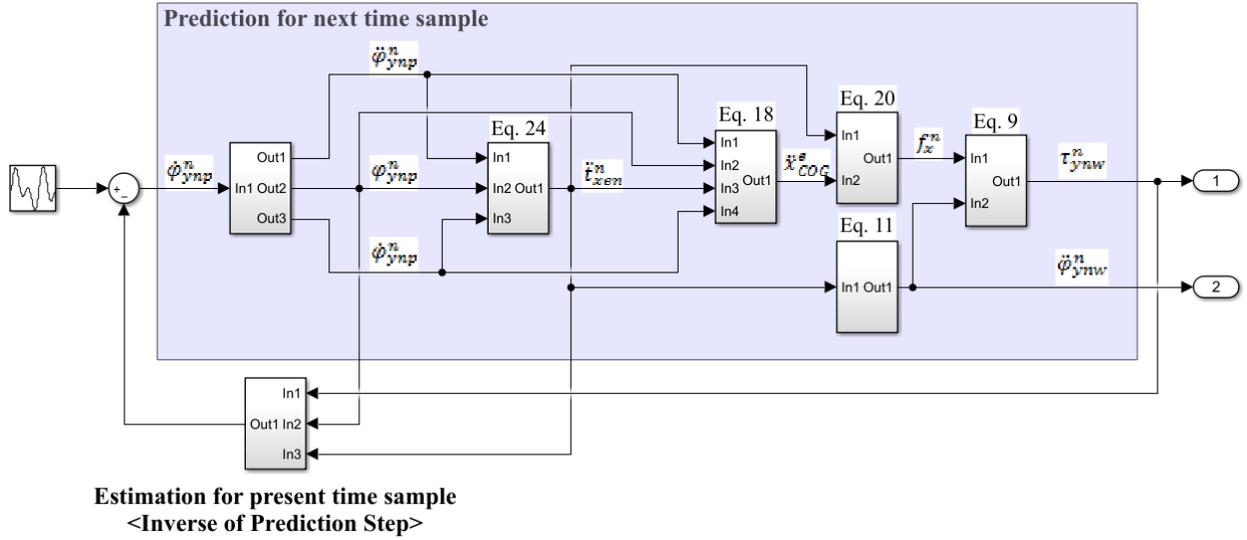

**Fig. 4:** MPC for a TWPTR

## 2-3.    TDC

The back electromotive force (EMF) or induced voltage of a DC motor, $e_m$, has a direct relationship with the magnetic field, $\phi_f$, and angular velocity of the shaft, $\dot{\theta}_m$. This relation is denoted by a constant factor $K_1$ in equation 25.

$$25.\ e_m = K_1 \phi_f \dot{\theta}_m$$

$\phi_f$ varies with the field current in a manner similar to the saturation characteristic of the magnetic material of the machine. It is a common practice to assume that the machine operates such that the magnetic field is proportional to the field current, $i_f$. As a result:

$$26.\ \begin{array}{l} \phi_f = K_2 i_f \\ K = K_1 K_2 \end{array} \rightarrow e_m = K i_f \dot{\theta}_m$$

Meanwhile, the generated torque of the motor, $\tau_m$, can be obtained from the basic relation $\frac{P_m}{\theta_m}$, where $P_m$ is the motor power. So the equation 27 can be concluded.

$$27.\ \tau_m = \frac{e_m i_m}{\dot{\theta}_m} \rightarrow \tau_m = K i_f i_m$$

In the motor circuit model, if the magnetic field is modeled as a series circuit, it can be concluded that $i_f$, $i_m$ and origin current, $i$, are equal. So the equation 28 is obtained.

$$28.\ \tau_m = K i^2 \rightarrow i = \pm\sqrt{\frac{\tau_m}{K}}$$

All DC motor equations are derived from [21]. The sign of $i$ is determined by the sign of $\varphi_{ynp}^n$ (If $\varphi_{ynp}^n$ is greater than zero, the sign will be plus, and vice versa). Also, the required torque to be generated by the DC motor of wheels, $\tau_{ynw}^n$, is determined by equation 23. If we assume $\tau_{ynw}^n$ is equal to $\tau_m$, the value of $i$ in each sample time will be determined.

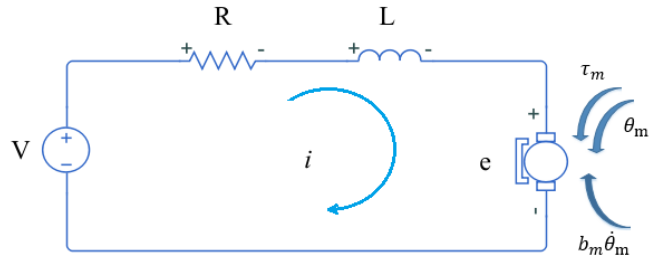

**Fig. 5:** Circuit model of DC motor

From Fig. 5, equation 29 can be derived based on Newton's 2$^{\text{nd}}$ law.

$$29.\ I_w \ddot{\theta}_m + b_m \dot{\theta}_m = \tau_m \rightarrow I_w \dddot{\theta}_m + b_m \ddot{\theta}_m = \dot{\tau}_m$$

$I_w$ is the inertia moment of each wheel, and $b_m$ is the motor viscous friction constant.

According to equation 11, and given the equality of $\theta_\mathrm{m}$ and $\varphi_{ynw}^n$, the value of $\ddot{\theta}_m$ in each sample time can be determined.

With regard to Fig. 5, equation 30 is derived from Kirchhoff's voltage law.

$$30.\ L\frac{di}{dt} + Ri + Ki\dot{\theta}_m = V$$

$R$ and $L$ are electric resistance and electric inductance of the motor, respectively. By replacing equation 28 in equation 30, the equation 31 is obtained.

$$31.\ L\frac{\dot{\tau}_m}{2\sqrt{K\tau_m}} + (R + K\dot{\theta}_m)\sqrt{\frac{\tau_m}{k}} = V$$

As with MPC step, the voltage required for the wheels' motor is determined (Fig. 6).

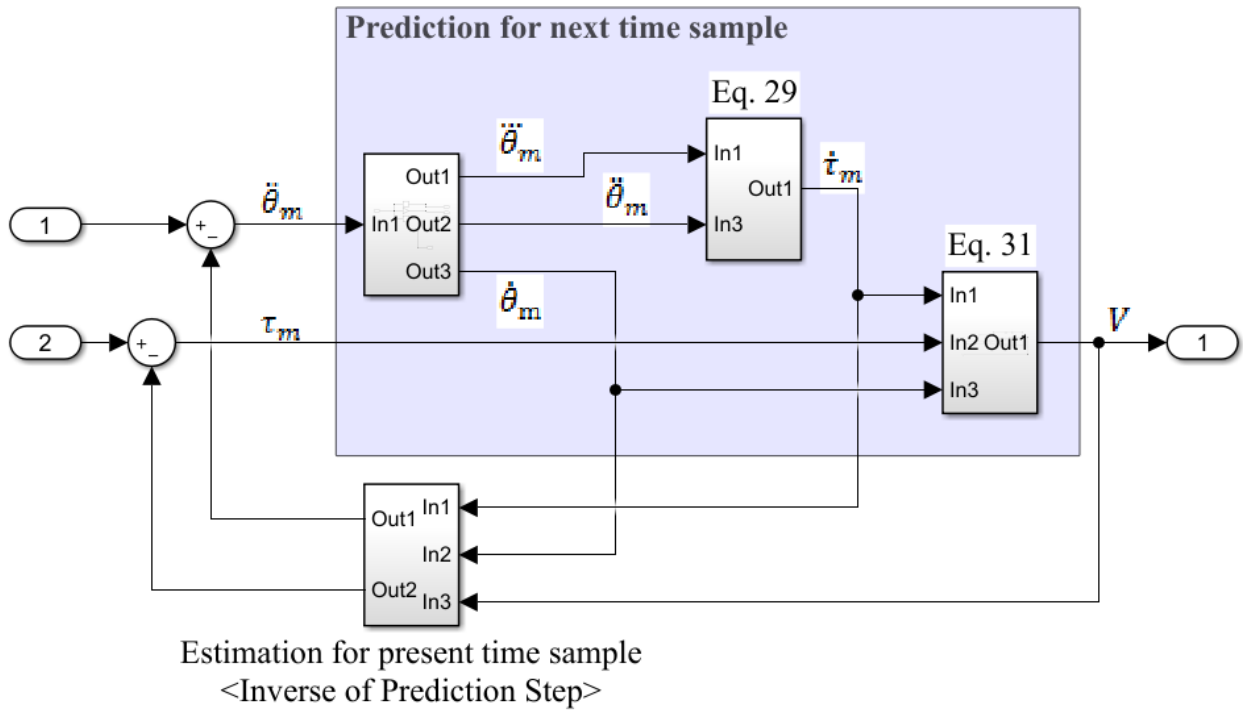

**Fig. 6:** TDC of a TWPTR

A detailed discussion on MPC can be found in [22].So, the hierarchical MPC method is the best approach to control the dynamics of the TWPTR and manage the disturbance of it simultaneously.

The block diagram of this method is shown in Fig. 7.

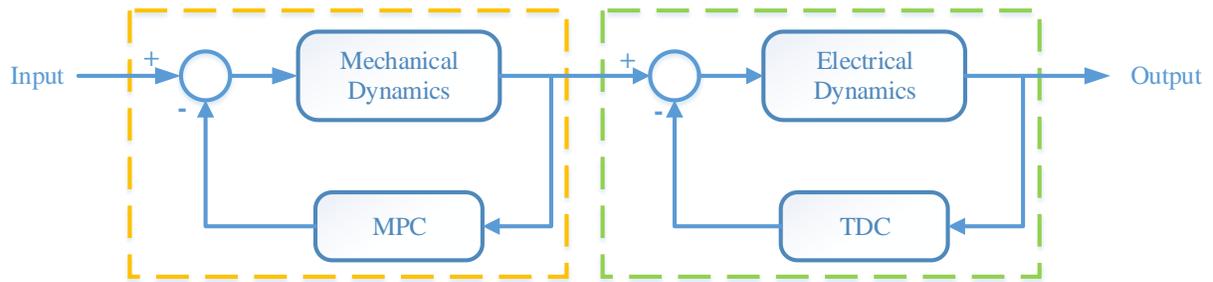

**Fig. 7:** Hierarchical MPC Block Diagram

## 2-4.    HIL Implementation

HIL simulation seeks to implement equations without any approximation. Reconstruct coefficients in the laboratory process are one of the advantages of this simulation [23]. In order to implement an HIL simulation, one should replace some virtual components in the simulation model with real components. The real components interact with the virtual components in real time to create an HIL simulation system [24]. Specifically, the output angular velocity of the pendulum as a real component is firstly measured and transmits data to the virtual component. Then, the response of different components including the output speed of the wheels' motor is observed (Fig. 8). For an HIL simulation test bench of a TWPTR, a real component will be a gyro connected to the pendulum. If the resulting output speed of DC motors can be fully tracked by controlling the angle of the pendulum, the virtual component is then well simulated.

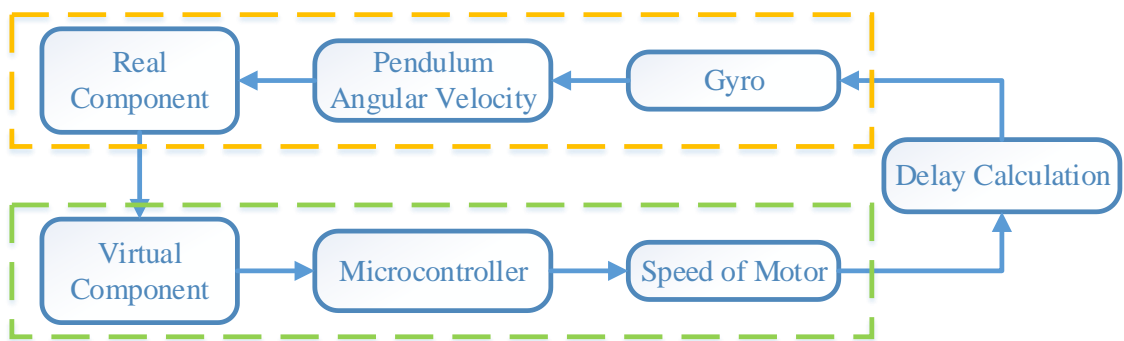

**Fig. 8:** HIL simulation for a TWPTR

## 3.   Simulation results

In this section, a TWPTR with defined dynamic parameters according to Table 1 is modeled.

First, the functionality of this robot with specific inputs and without any controller has been investigated, then the efficiency of the MPC algorithm has been evaluated. After that, the MPC has been evaluated against turbulence and delayed response of the robot engines. In the following, the combination of MPC and TDC to manage disturbances has been evaluated. In the end, to verify results the HIL simulation is utilized.

**Tab. 1:** TWPTR's parameters

| Parameters | Symbol | Value |
|---|---|---|
| Wheel's radius | $r$ | 20 cm |
| Length of COG | $l$ | 1 m |
| Mass of wheels | $m$ | 4 kg |
| Mass of the robot minus the wheels | $M$ | 100 kg |
| Moments of wheels | $I_w$ | $0.07 kgm^2$ |
| Robot pendulum moment | $I_p$ | $86.67 kgm^2$ |
| Earth's gravity acceleration | $g$ | $10 \dfrac{m}{s^2}$ |
| Transitional torque losses | $\tau_{fwp}$ | . |
| Gyro sampling frequency | - | 100 Hz |

It is assumed that a person on the TWPTR will do arbitrary moves to test the stability of the robot. In order to model the person's movements, an equation is considered as follows.

$$32. \; \dot{\phi}_{ynp}^n = 5\, sin(50t) + 4\, cos(20t).$$

Two points in equation 32 must be considered. First, due to the robot's structure, the person's maneuverability is low for rotation around the $y^n$; so the coefficients of the terms in equation 25 are set to be larger than the values that occur in reality. In other words, with regard to this equation, the maximum rotation of a person aroud the y-axis is 120 °, while the maximum rotation in reality is far less than this value. Secondly, it is assumed that a person has a significant COG position change, so the frequency of terms in equation 32 has a large value. In other words, this significant change is considered for testing the stability of the applied control method. If no control is applied to the system, the angle of the p-frame with respect to the n-frame can be shown in Fig. 9.

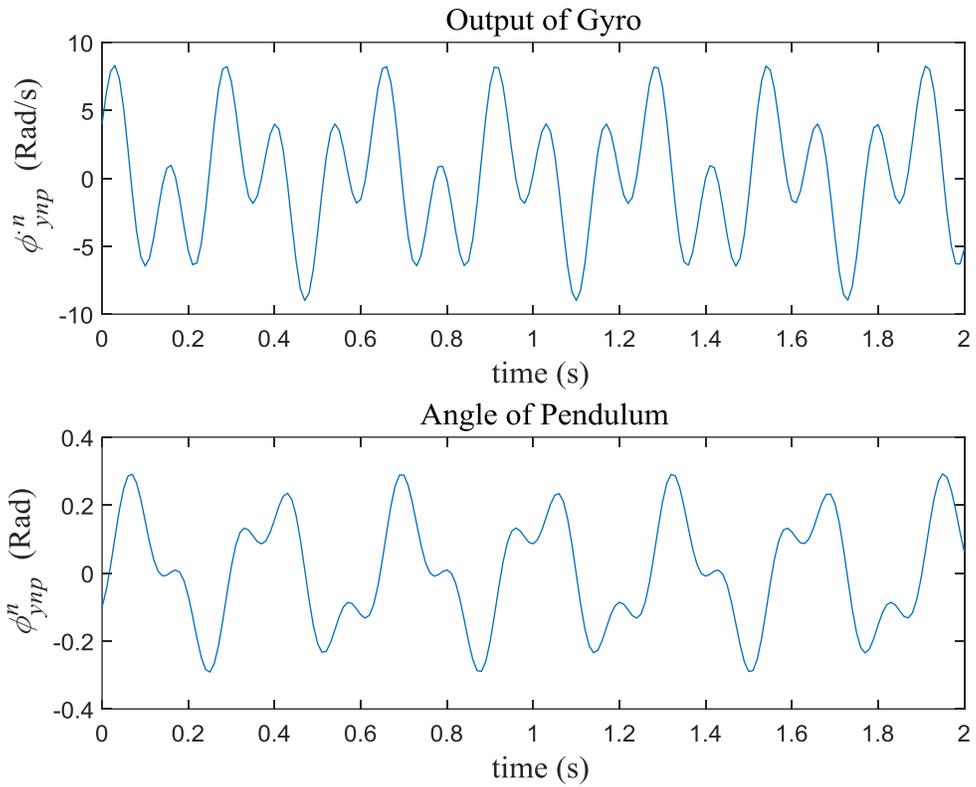

**Fig. 9:** The angular velocity and angular position of the p-frame with respect to the n-frame without any controller

After implementing the MPC proposed in Section 2-2, the output waveform can be seen in Fig. 10. Indeed, this shape is related to the angle of the p-frame around the $y^n$. The minor changes in the position of the pendulum in Fig. 10 is because of the assumption that the angular acceleration in the previous and present sample time is considered equal. Also, this figure also includes the torque input to the motors over time.

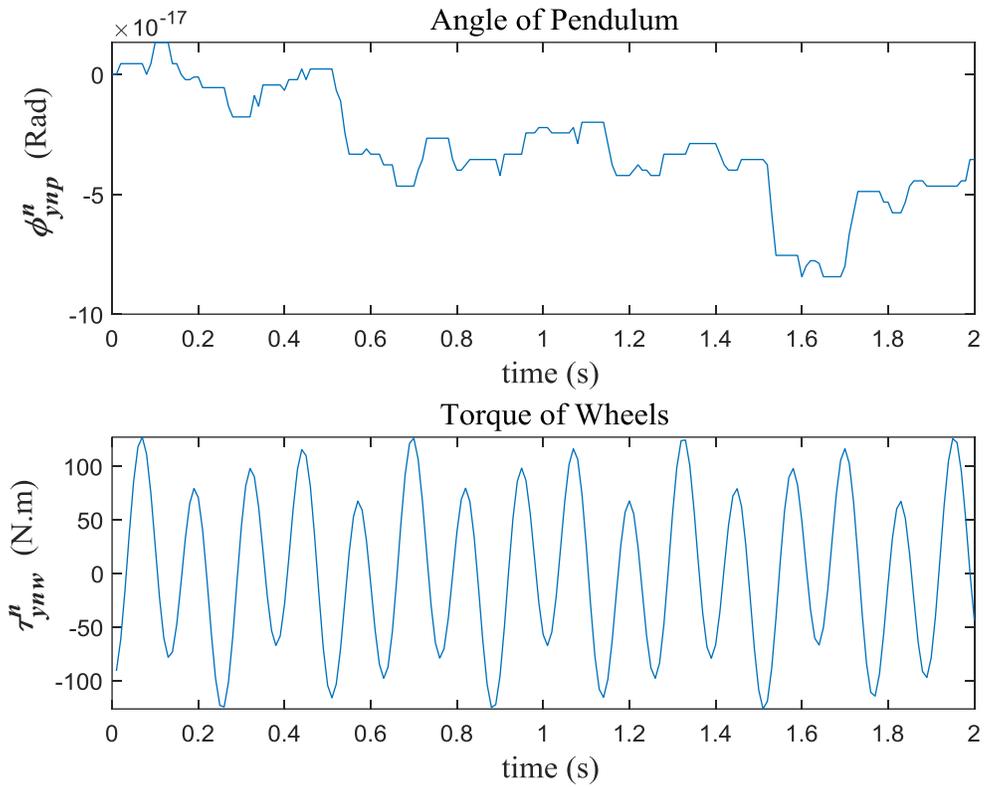

**Fig. 10:** The angular position of the p-frame with respect to the n-frame without considering the delay response of wheels' motor – the torque of each wheels' motor

In Fig. 10, is assumed that the DC motors generate the required torque in real time. In a real robot, such an assumption is far from the reality, so Fig. 11 displays the pendulum angle, considering the two sampling units for delay of motor response. By comparing Fig. 10 and Fig. 11, it can be concluded that consideration of the delay response caused a kind of nonstablization in pendulum angular position. Also, the torque required for its control grows as demonstrated in Fig. 12. Therefore, implementing the TDC is important to control the TWPTR and decrease the maximum required torque.

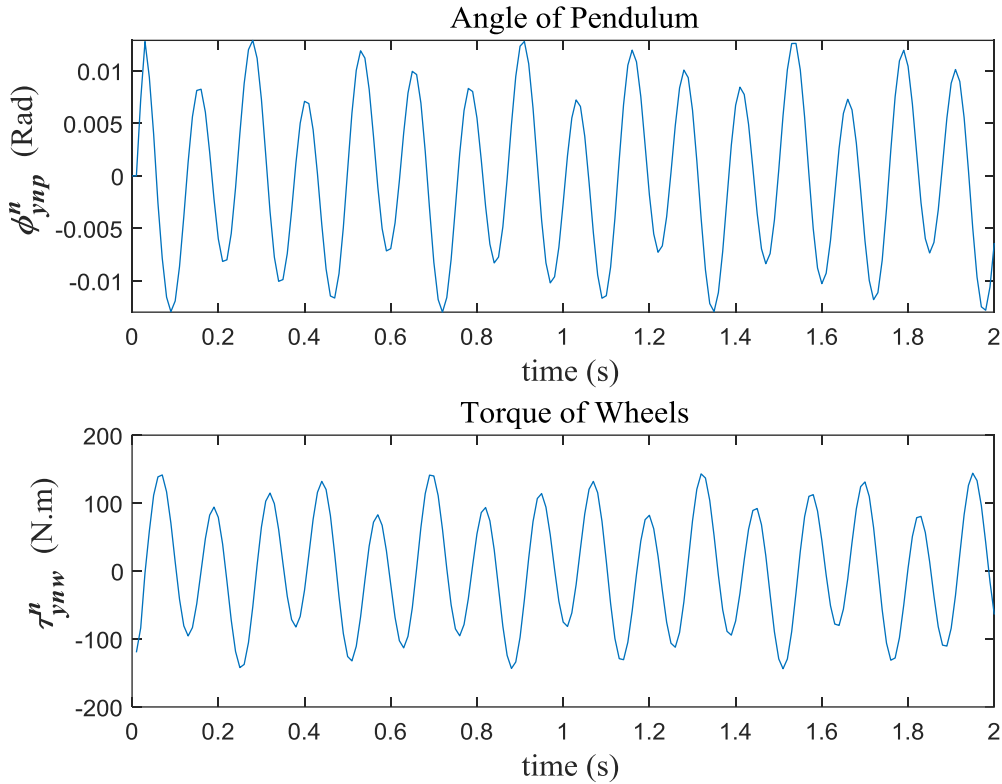

**Fig. 11:** The angular position of the p-frame with respect to the n-frame considering the delay response of wheels' motor – the torque of each wheels' motor

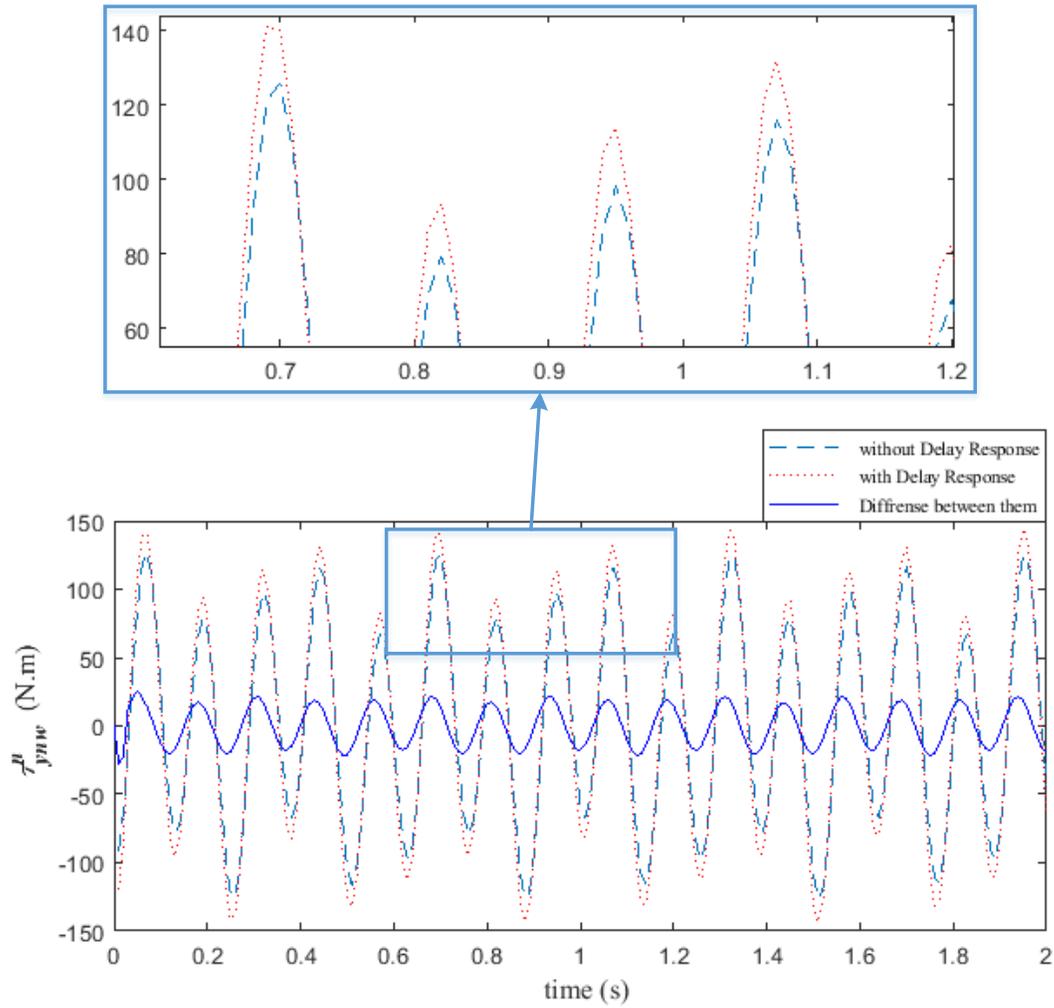

**Fig. 12:** Torque of each wheels' motor with delay and nondelay response and difference between them

After implementing the TDC proposed in Section 2-3, the pendulum angular position can be seen in Fig. 13. By comparing Fig. 11 and Fig. 13, it can be concluded that applying the TDC has caused a significant reduction in the angular movement of the pendulum, or rather the maximum displacement of the robot around the balance axis has decreased 44%. Precisely, robot stability has increased by 78%. Also, the torque required for its control decreases as indicated in Fig. 14.

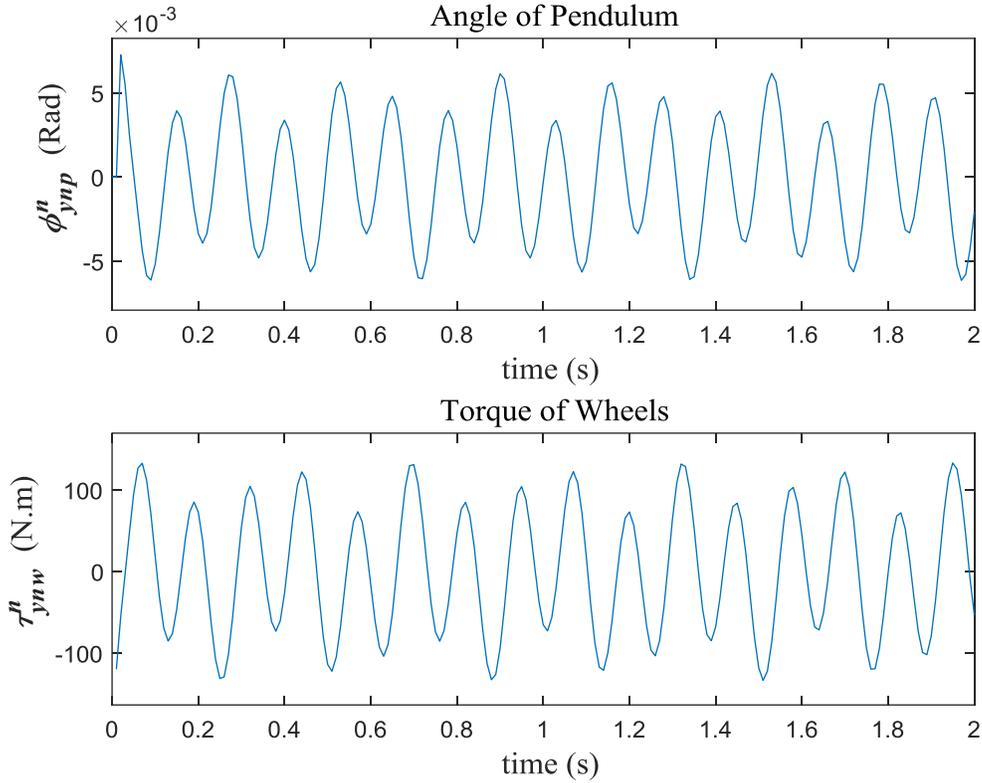

**Fig. 13:** The angular position of the p-frame with respect to the n-frame by applying TDC - torque of each wheels' motor

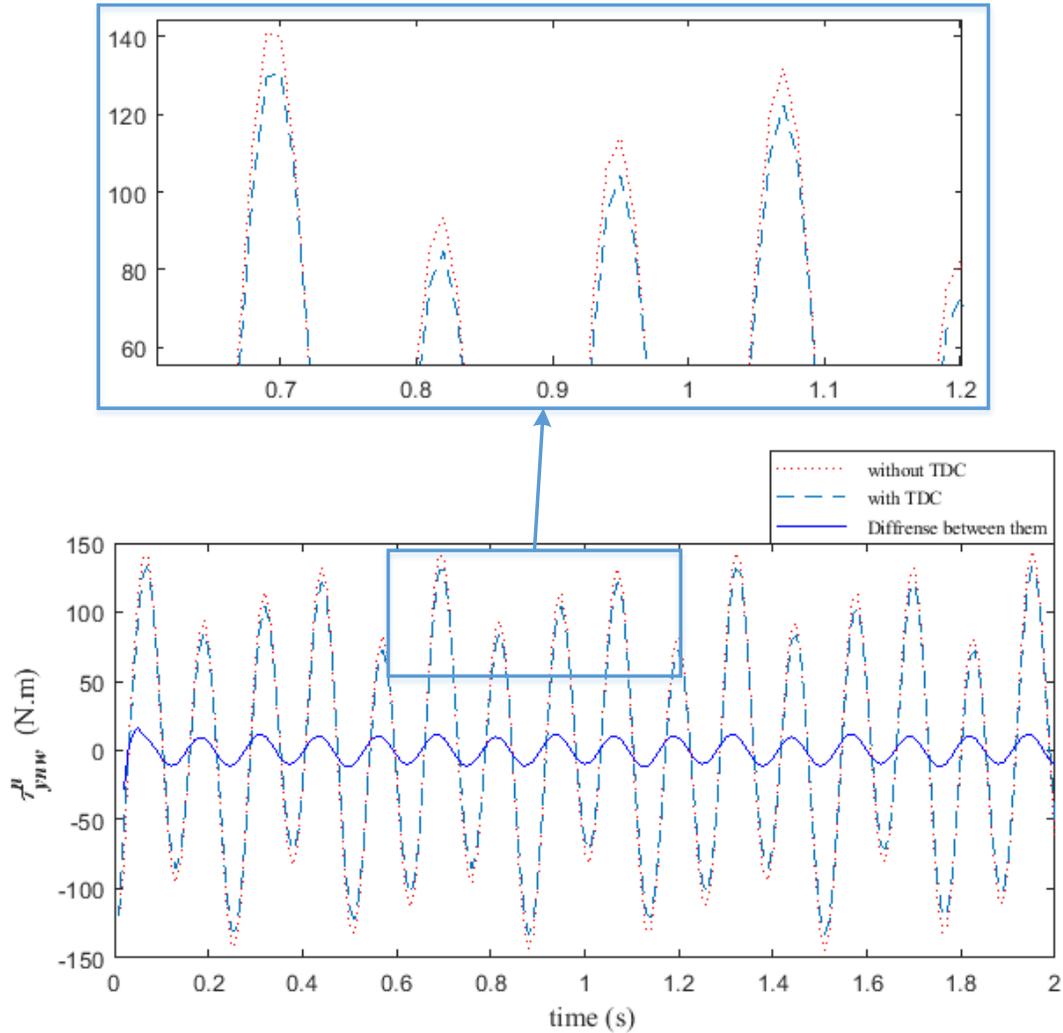

**Fig. 14:** Torque of each wheels' motor with TDC and without TDC and the difference between them

According to Fig. 14, when TDC method is applied to output of MPC the results show a 7% reduction in the maximum torque required by the engines.

In order to verify the performance of the hierarchical MPC method, different tests are implemented in this section. Firstly, a start-up procedure of TWPTR controller hardware is described step by step. Then, the motor speed delay time response is investigated. A comparison is provided on the responses of the TWPTR in the HIL simulation and the analytical simulation.

To start up the system, real-time simulation is first implemented in PROTEUS. To apply this simulation, all electric elements are precisely connected together according to real controller electrical circuit of a TWPTR. Thereafter, the gyro measurement data are sent to the microcontroller. With the initiation of the control process, modeled in the HIL platform, two virtual DC motors are turned and accelerated to the desired speed. When these motors reach the desired speed, the electrical voltage of the motors should be sent from the microcontroller. Because of PWM control of motor speed, we use encoder motor in this simulation and send their real-time speed to the microcontroller at each sample time. When the speed signal of the DC motors hits the set point, a digital signal is sent to the gyro sensor. The microcontroller checks the synchronization by checking the speed and phase differences of the gyro sensor and DC motors, as revealed in Fig. 15.

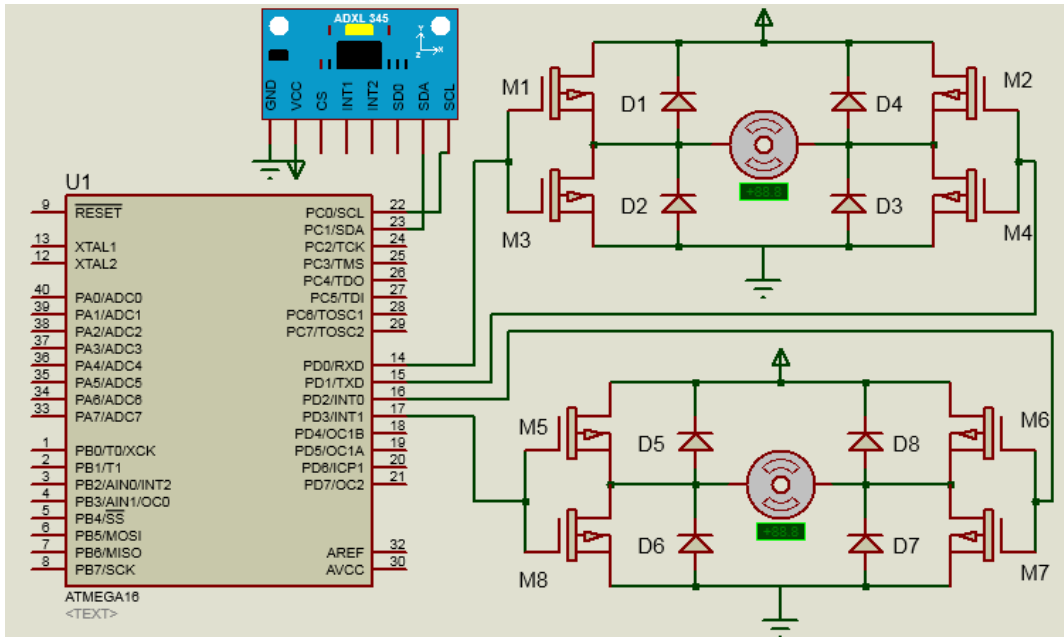

**Fig. 15:** Electrical Circuit Simulation of a TWPTR

## 4. Conclusion

According to the simulation results and based on the maximum pendulum angular position as well as the maximum torque applied to the motors, the system's response to different inputs can be stabilized rapidly and accurately. Although this robot has non-negligible nonlinearities, hierarchical MPC method control TWPTR considers all these nonlinearities while conserving the simplicity of implementation. Also, according to the HIL simulation results, the proper structure can be designed for manufacturing robots. The need for HIL simulation becomes clearer when changing the inherent properties of TWPTR in the control process becomes necessary.

In two-wheeled autonomous robots, determining the correct position is essential. The inertial measurement unit (IMU) is generally used for higher precision to determine robot position. Measuring the rotation matrix between the IMU-frame and coordinates systems of the two-wheeled robot is necessary for robot control process. This issue can be an interesting topic for further research.

   The other interesting topic for research is the segway's field pressure control. Due to the structure of these robots, the extent of force that pulls the passenger during the robot movements is very important. In addition to its direct impact on the safety, by minimizing this force, the magnitude of required torque applied to the wheel's motor can be decreased. Eventually, by reducing the maximum torque, the cost of the robot will also be reduced.

**"Conflict of interest – none declared"**